\def\bar{\begin{eqnarray}}
\def\ear{\end{eqnarray}}

\def\eqi{\begin{equation}}
\def\eqf{\end{equation}}
\def\eqia{\begin{eqnarray}}
\def\eqfa{\end{eqnarray}}
\def\rp#1#2{{#1\over#2}}
\def\ct#1{\cite{#1}}

\def\oc2{$\mathcal{O}(c^{-2})$}

\documentclass[11pt]{article}
\usepackage{amsmath,amsthm,amscd,amssymb}
\usepackage{latexsym}
\usepackage{graphicx,epsfig}

\begin{document}

\noindent{\bf \LARGE{The impact of the new Earth gravity model
EIGEN-CG03C on the measurement of the Lense-Thirring effect with
some existing Earth satellites }}
\\
\\
\\
{Lorenzo Iorio, {\it FRAS, DDG}}\\
{\it
\\Viale Unit$\grave{a}$ di Italia 68, 70125\\Bari, Italy
\\e-mail: lorenzo.iorio@libero.it}

\begin{abstract}
The impact of the latest combined CHAMP/GRACE/terrestrial
measurements Earth gravity model EIGEN-CG03C on the measurement of
the Lense-Thirring effect with some linear combinations of the
nodes of some of the existing Earth's artificial satellites is
presented. The 1-sigma upper bound of the systematic error in the
node-node LAGEOS-LAGEOS II combination is $3.9\%$ (4$\%$ with
EIGEN-GRACE02S, $\sim 6\%$ with EIGEN-CG01C and $\sim 9\%$ with
GGM02S), while it is 1$\%$ for the node-only LAGEOS-LAGEOS
II-Ajisai-Jason-1 combination (2$\%$ with EIGEN-GRACE02S, $ 1.6\%$
with EIGEN-CG01C and $2.7\%$ with GGM02S).
\end{abstract}
\section{Introduction}
In this brief note we release an update about the impact of the
Newtonian part of the terrestrial gravitational potential on the
measurement of the general relativistic gravitomagentic
Lense-Thirring effect \ct{leti} with some linear combinations
\ct{ciuf96, iormor, iorMGM, iorproc, nova, ves, iordorn} of the
orbital residuals of the nodes $\Omega$ and the perigees $\omega$
of certain existing Earth's artificial satellites. The Earth
gravity model EIGEN-CG03C \ct{cg03}, recently released by the
GeoForschungsZentrum\footnote{See on the WEB
http://www.gfz-potsdam.de/pb1/op/grace/results/index$\_$RESULTS.html
for retrieving the model} (GFZ), is considered here.

The gravity field combination model EIGEN-CG03C is an upgrade of
EIGEN-CG01C \ct{cg01}. The model is based on the same CHAMP
mission and surface data (0.5 x 0.5 deg. gravimetry and
altimetry), but takes into account almost twice as much GRACE
mission data. Instead of 200 days now 376 days out of February to
May 2003, July to December 2003 and February to July 2004 have
been used.

Let us recall that one of the most important sources of systematic
error in measuring the Lense-Thirring precessions with the data
from the Earth's satellites is represented by the much larger
classical effects induced by the static and time--varying parts of
the even zonal harmonic coefficients $J_{\ell},\ \dot J_{\ell}\
\ell=2,4,6,...$ of the multipolar expansion of the terrestrial
gravitational potential. Indeed, the accuracy with which such
parameters are currently known does not yet allow to analyze only
one satellite. The linear combination approach, proposed for the
first time by Ciufolini \ct{ciuf96} and, subsequently, extended
and generalized by Iorio \ct{iormor, iorMGM, iorproc, nova,
iordorn}, consists in suitably combining the orbital elements of
different satellites in order to disentangle, by construction, the
Lense-Thirring effect from the largest number of even zonals as
possible.
\section{Some linear combinations}
\subsection{The node-node-perigee combination with the LAGEOS satellites}
The first combination proposed and analyzed with some of the
latest pre-CHAMP/GRACE Earth gravity models involves the nodes of
LAGEOS and LAGEOS II and the perigee of LAGEOS II \ct{ciuf96}. It
cancels out $J_2$ and $J_4$ but is heavily affected by the
non-gravitational perturbations (direct solar radiation pressure,
Earth's albedo, thermal effects as the solar Yarkovsky-Schach and
the terrestrial Yarkovsky-Rubincam forces) which act on the
perigee of the LAGEOS-type satellites.
\subsection{The node-node combination with the LAGEOS satellites}
Later, the improvements in our knowledge of the terrestrial
gravitational field due to the first CHAMP and, especially, GRACE
models allowed to explore different alternatives. The most viable,
from the point of view of the reduction of the non-gravitational
perturbations and of the relative easiness of  the data-analyzing
process, is the $J_2-$free combination\footnote{The idea of only
using the nodes of the LAGEOS satellites in view of the  results
from the GRACE mission was qualitatively put forth for the first
time in \ct{ries}.} of the nodes of LAGEOS and LAGEOS II
\ct{iormor, iorMGM, iorproc, nova} which has recently been used in
a test with real data over 11 years \ct{nat}. It has the great
advantage of discarding the perigee of LAGEOS II. However, it is
affected by $J_4, J_6, J_8...,\dot J_4,\ \dot J_6$ whose impact on
the total available accuracy is still rather controversial
\ct{NA1, NA2}. It is so mainly because the error due to the static
part of the geopotential even zonals $J_4, J_6, J_8,...$ is still
rather model-dependent ranging from $\sim 4\%$ to $\sim 9\%$ at
1-sigma level. Moreover,  the effect of the secular rates $\dot
J_4, \dot J_6$, which could induce an additional $\sim 11\%$ bias
of the Lense-Thirring effect thus increasing the total error to
19-24$\%$ at 1-sigma level, is important as well.
\subsection{The node-only LAGEOS-LAGEOS II-Ajisai-Jason-1 combination}
A combination involving the nodes of LAGEOS, LAGEOS II, Ajisai and
the altimeter satellite Jason-1 has been proposed in \ct{iordorn,
ves}. It cancels out $J_2,J_4,J_6$ along with their secular
variations $\dot J_2,\dot J_4,\dot J_6$. As a consequence, the
systematic error due to the remaining even zonal harmonics is
smaller ($\sim 1-2\%$) and less model-dependent than that of the
node-node LAGEOS-LAGEOS II combination. Another important point is
that GRACE should better improve the mid-high degree even zonal
harmonics which are just the most relevant for such a combination.
Thus, it is likely that  the error of gravitational origin will be
further reduced well below the 1$\%$ level with the forthcoming,
more robust solutions from GRACE. Instead, it might be not so for
the node-node LAGEOS-LAGEOS II combination because of the fact
that it is affected by the low-degree zonals $J_4,\ J_6$ for which
the future improvements by GRACE should be less relevant. The
weakest point of the use of Jason-1 is represented by the
non-gravitational perturbations acting on such a spacecraft of
complex shape and attitude. Indeed, the Jason's area-to-mass ratio
$S/M$, to which the non-gravitational forces are proportional, is
almost two orders of magnitude larger than that of the LAGEOS
satellites. On the other hand, the coefficient with which Jason-1
enters the combination is 0.068, i.e., just two orders of
magnitude smaller than the coefficients weighing LAGEOS (1) and
LAGEOS II (0.347). Moreover, no secular or sinusoidal
perturbations with long periods should affect the node of Jason-1.
Finally, the orbital maneuvers which periodically are performed
are mainly, although not entirely, in-plane, while the node is
related to the out-of-plane component of the orbital path. Thus,
the implementation of such a combination, although certainly more
difficult than the analysis of the observables involving LAGEOS
and LAGEOS II, should deserve attention from the geodesist's
community.
\subsection{The node-only LAGEOS-LAGEOS II-Ajisai-Starlette-Stella combination}
A combination involving only the nodes of the currently best
tracked geodetic satellites LAGEOS, LAGEOS II, Ajisai, Starlette
and Stella was also proposed \ct{iormor, iorMGM}. However, it
turned out to be not competitive with the other implemented or
proposed combinations because of the larger number of even zonal
harmonics of high degree to which the other low-altitude
satellites like Starlette and Stella are sensistive.
\section{The results}
The results obtained with EIGEN-CG03C are shown in Table
\ref{tavola}.
\begin{table}
\caption{Percent systematic error
$\left.\rp{\delta\mu}{\mu}\right|_{\rm even\ zonals }$ in the
measurement of the Lense-Thirring effect with various linear
combinations of the nodes $\Omega$ and the perigees $\omega$ of
some existing geodetic and altimeter satellites due to the
uncancelled even zonal harmonics of the geopotential according to
the calibrated sigmas of the EIGEN-CG03C Earth gravity model. The
quoted errors are 1-sigma level upper bounds calculated by
linearly adding over degree $\ell$ the absolute values of the
combined mismodelled precessions. No covariance matrix was used.
The 1-sigma root-sum-square errors are quoted in round brackets.
L=LAGEOS, L II=LAGEOS II, Aji=Ajisai, Jas=Jason-1, Str=Starlette,
Stl=Stella. }\label{tavola}
\begin{tabular}{@{\hspace{0pt}}ll}
\hline\noalign{\smallskip} Combination & $\rp{\delta\mu}{\mu}$
($\%$)
\\
\noalign{\smallskip}\hline\noalign{\smallskip} $\Omega^{\rm
L}-\Omega^{\rm L\ II}-\omega^{\rm L\ II}$ \ct{ciuf96}& 0.6 (0.4)\\
$\Omega^{\rm L}-\Omega^{\rm L\ II}$ \ct{iormor, iorMGM, iorproc, nova}& 3.9 (3.0)\\
$\Omega^{\rm L}-\Omega^{\rm L\ II}-\Omega^{\rm Aji}-\Omega^{\rm Jas}$  \ct{iordorn, ves} & 1.0 (0.5)\\
$\Omega^{\rm L}-\Omega^{\rm L\ II}-\Omega^{\rm Aji}-\Omega^{\rm Str}-\Omega^{\rm Stl}$ \ct{iormor, iorMGM, ves}& 19.9 (6.2)\\
 \noalign{\smallskip}\hline
\end{tabular}
\end{table}
It should be noted that the 1-sigma upper bounds of the error in
the node-node LAGEOS-LAGEOS II combination are $\sim 6\%$ with
EIGEN-CG01C, $4\%$ with the GRACE-only GFZ EIGEN-GRACE02S solution
\ct{grace02s} and $\sim 9\%$ with the Center for Space Research
(CSR/UT) GRACE-only GGM02S model \ct{ggm02}. In regard to the
combination involving the nodes of LAGEOS, LAGEOS II, Ajisai and
Jason-1, EIGEN-CG01C yields $1.6\%$, EIGEN-GRACE02S $2\%$ and
GGM02S $2.7\%$. The combination of the nodes of LAGEOS, LAGEOS II,
Ajisai, Starlette and Stella still remains far from the achievable
levels obtainable with the combination of the nodes of LAGEOS and
LAGEOS II  and the combination of the nodes of LAGEOS, LAGEOS II,
Ajisai and Jason-1.


\end{document}